\documentstyle[preprint,eclepsf]{jpsj}

\title{
One-Particle Excitation\\
of the Two-Dimensional Hubbard Model\footnote{
To appear in J. Phys. Soc. Jpn. {\bf 66} No. 3 (1997).} 
}

\author{
Ken {\sc Yokoyama}\footnote{E-mail: yokoyama@watson.phys.s.u-tokyo.ac.jp}
and Hidetoshi {\sc Fukuyama}
}

\inst{
Department of Physics, Tokyo University, Tokyo 113
}

\recdate{January 21, 1997}

\abst{
The real part of the self-energy of interacting two-dimensional electrons
has been calculated in the {\em t}-matrix approximation.
It is shown that the forward scattering results in an anomalous term
leading to the vanishing renormalization factor of the one-particle
Green function,
which is a non-perturbative effect of the interaction $U$.
The present result is a microscopic demonstration of the claim by
Anderson based on the conventional many-body theory.
The effect of the damping of the interacting electrons,
which has been ignored in reaching above conclusion, has been briefly
discussed.
}

\kword{
Hubbard model, two dimensions, {\em t}-matrix approximation, forward scattering
}

\begin{document}
\sloppy
\maketitle

\newcommand{\rRe}{{\rm Re}}
\newcommand{\iIm}{{\rm Im}}
\newcommand{\sign}{{\rm sgn}}
 
The nature of the low-energy excitation of the two-dimensional system
is of great interest recently.
The Fermi-liquid picture was considered to be valid from diagramatic
studies,\cite{bib:Hodges} while it was suggested by Anderson
that the anomalous behavior of the forward scattering
phase shift leads to the renormalization factor $Z=0$, i.e.,
the breakdown of the Fermi-liquid.\cite{bib:Anderson1,bib:Anderson2}
This remarkable suggestion has attracted much interest,
\cite{bib:Fab,bib:Stamp}
and several calculations of the self-energy
of the two-dimensional Hubbard model have been carried out
based on the {\em t}-matrix approximation,
\cite{bib:Engelbrecht,bib:FH,bib:FHN,bib:Narikiyo}
in which the self-energy is 
approximated by the summation of ladder diagrams of the particle-particle
process.
In these calculations, however, only the imaginary part of the self-energy 
has been considered, and the real part has not been studied in detail.

In this paper, we will calculate explicitly the real part of 
the self-energy of the two-dimensional Hubbard model by the {\em t}-matrix
approximation,  and show that the
singularity of the {\em t}-matrix in the forward scattering region gives rise to
an anomalous term to the real part of the self-energy,
which leads to the renormalization factor $Z=0$.
This result is in accordance with the claim by Anderson.

We consider the asymptotic behavior of the self-energy 
$\Sigma(\mib{k},\epsilon+{\rm i}\delta)$ in the limit of
$|\mib{k}-\mib{k}_{\rm F}|\ll k_{\rm F}, |\epsilon| \ll \epsilon_{\rm F}$,
$\mib{k}_{\rm F}$ and $\epsilon_{\rm F}$ being the Fermi momentum and the Fermi energy. 
In general, the shape of the Fermi 
surface near the point $\mib{k}$ in the momentum space
can be approximated as a parabolic 
curve, as is schematically shown in Fig. \ref{fig:dispersion}(a) and
\ref{fig:dispersion}(b).
Linearizing the energy dispersion in the normal 
direction of the Fermi surface, we assume the following 
energy dispersion.\cite{bib:Ogata}
\begin{equation}
        \xi_k\equiv\epsilon_k-\mu\equiv v_0k_x+\frac{A_0}{2} k_y^2,
    \label{eq:dispersion}
\end{equation}
where we take the origin of momentum at the point nearest to
$\mib{k}$ on the Fermi surface, as is shown in Fig. \ref{fig:dispersion}(b),
$\mu$, $v_0$ and $A_0$ are the chemical potential, the Fermi-velocity and
a constant, respectively.
Here, the momentum dependence of the velocity is neglected.
\begin{fullfigure}
         \epsfile{file=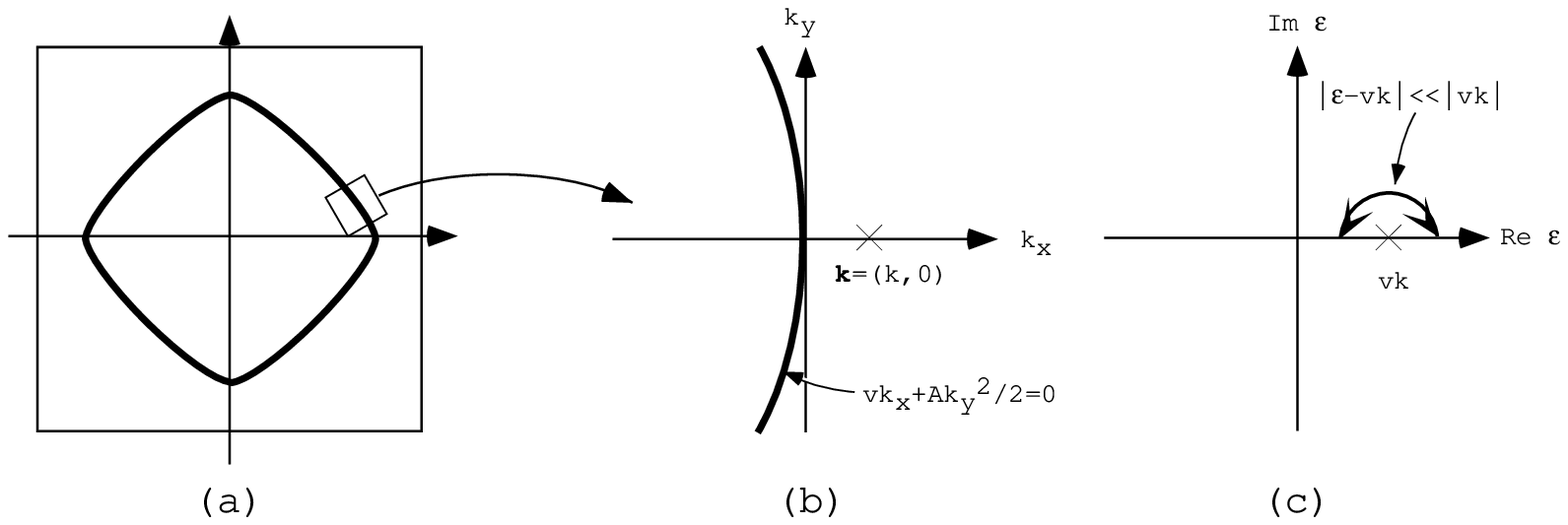,height=4cm}
  \caption{The local shape of (a) the Fermi surface can be approximated by
    (b) a parabolic curve. 
    (c) The energy $\epsilon$ dependence of the self-energy
    in the vicinity of $vk$ is focused on.}
  \label{fig:dispersion}
\end{fullfigure}
We calculate the self-energy
$\Sigma(\mib{k},\epsilon+{\rm i}\delta)$ at the point $\mib{k}=(k,0)$.
In the Hubbard model, the {\em t}-matrix is given by
\begin{eqnarray}
  T(\mib{q},x+{\rm i}\delta)&=&\frac{-U}{1+UK(\mib{q},x+{\rm i}\delta)}\\
  K(\mib{q},x+{\rm i}\delta)&=&\sum_{k}\frac{\sign\xi_{k+q/2}}
    {\xi_{k+q/2}+\xi_{-k+q/2}-x-{\rm i}\delta},
  \label{eq:pp_cor}
\end{eqnarray}
where $K(\mib{q},x+{\rm i}\delta)$ is the particle-particle correlation 
function. Noting that
$K(\mib{q},0)$ as a function of $\mib{q}$ is regular in the
forward scattering region, $\mib{q}\simeq 2\mib{k}_{\rm F}$,
we introduce
$K_{0}\equiv\lim_{q\rightarrow 2k_{\rm F}}\left[
\lim_{x\rightarrow 0} K(\mib{q},x+{\rm i}\delta)\right]$,
which has a contribution from the high-energy region, and
reflects the whole band structure.
In contrast to this, $q$- and $x$-dependences of
$K(\mib{q},x+{\rm i}\delta)-K_{0}$ in the region of
$|\mib{q}-2\mib{k}_{\rm F}|\ll k_{\rm F}, |x| \ll \epsilon_{\rm F}$ 
reflect details of a scattering process near the Fermi-energy.
Hence, we write the {\em t}-matrix as,
\begin{eqnarray}
  T(\mib{q},x+{\rm i}\delta)&=&\frac{-U_{\rm eff}}
    {1+U_{\rm eff}\left[K(\mib{q},x+{\rm i}\delta)-K_{0}\right]}\\
  U_{\rm eff}&\equiv&\frac{U}{1+UK_{0}}.
\end{eqnarray}
We will focus only on the contribution of the {\em t}-matrix
to $\Sigma(\mib{k},\epsilon+{\rm i}\delta)$
from the forward scattering region, i.e., the total momentum of the 
particle-particle ladder is nearly $2\mib{k}$.
In the following calculations, we introduce momentum cut-offs
$k_{\rm c}$ and $k_{\rm c}'$ for $x$ and $y$ components of the momentum
in the energy dispersion, and momentum integrations for intermediate
states of a scattering process are carried out within these cut-offs.

In the {\em t}-matrix approximation, the expressions for the real and
imaginary parts of the self-energy are given by
\begin{eqnarray}
  \rRe\Sigma ({\mib k},\epsilon+{\rm i}\delta)&=&
           \int\frac{{\rm d}^2q}{(2\pi)^2}
                \int_{\epsilon}^{\infty}\frac{{\rm d}x}{\pi}
    \rRe T({\mib q},x+{\rm i}\delta)
    \iIm G_0({\mib q}-{\mib k},x-\epsilon-{\rm i}\delta)
    \nonumber\\
  & &-\int\frac{{\rm d}^2q}{(2\pi)^2}
    \int_{0}^{\infty}\frac{{\rm d}x}{\pi}    
    \iIm T({\mib q},x+{\rm i}\delta)
    \rRe G_0({\mib q}-{\mib k},x-\epsilon-{\rm i}\delta)
    \label{eq:res}\\
  \iIm\Sigma ({\mib k},\epsilon+{\rm i}\delta)&=&
           -\int\frac{{\rm d}^2q}{(2\pi)^2}
    \int_{0}^{\epsilon}\frac{{\rm d}x}{\pi}
    \iIm T({\mib q},x+{\rm i}\delta)
    \iIm G_0({\mib q}-{\mib k},x-\epsilon-{\rm i}\delta).
    \label{eq:ims}
\end{eqnarray}
We will evaluate the asymptotic forms of
$\rRe\Sigma ({\mib k},\epsilon+{\rm i}\delta)$ and
$\iIm\Sigma ({\mib k},\epsilon+{\rm i}\delta)$ under the condition of
$|\epsilon|,|vk|\ll \epsilon_{\rm c}$, where $\epsilon_{\rm c}$ is defined as
${\rm Min}(v_0k_{\rm c}, A_0k_{\rm c}'^2)$.
Dividing $x$ integration in $[\epsilon,\infty]$ in
the first term of eq. (\ref{eq:res})
into $[0,\epsilon]$ and $[0,\infty]$,
we introduce $\rRe\Sigma_{1}({\mib k},\epsilon+{\rm i}\delta)$,
$\rRe\Sigma_{2}({\mib k},\epsilon+{\rm i}\delta)$ and
$\rRe\Sigma_{3}({\mib k},\epsilon+{\rm i}\delta)$ as
\begin{eqnarray}
  \rRe\Sigma({\mib k},\epsilon+{\rm i}\delta)&\equiv&
    \rRe\Sigma_{1}({\mib k},\epsilon+{\rm i}\delta)
    +\rRe\Sigma_{2}({\mib k},\epsilon+{\rm i}\delta)
    +\rRe\Sigma_{3}({\mib k},\epsilon+{\rm i}\delta)\\
  \rRe\Sigma_{1}({\mib k},\epsilon+{\rm i}\delta)&\equiv&
    \int\frac{{\rm d}^2q}{(2\pi)^2}
                \int_{0}^{\infty}\frac{{\rm d}x}{\pi}\left[
    \rRe T({\mib q},x+{\rm i}\delta)
    \iIm G_0({\mib q}-{\mib k},x-\epsilon-{\rm i}\delta)
    \right.\nonumber\\
  & &\left.-\iIm T({\mib q},x+{\rm i}\delta)
    \rRe G_0({\mib q}-{\mib k},x-\epsilon-{\rm i}\delta)\right]\\
  \rRe\Sigma_{2}({\mib k},\epsilon+{\rm i}\delta)&\equiv&
           -\int\frac{{\rm d}^2q}{(2\pi)^2}
    \int_{0}^{\epsilon}\frac{{\rm d}x}{\pi}
    \rRe T({\mib q},0)\iIm G_0({\mib q}-{\mib k},x-\epsilon-{\rm i}\delta)\\
  \rRe\Sigma_{3}({\mib k},\epsilon+{\rm i}\delta)&\equiv&
           -\int\frac{{\rm d}^2q}{(2\pi)^2}
    \int_{0}^{\epsilon}\frac{{\rm d}x}{\pi}
    \left[\rRe T({\mib q},x+{\rm i}\delta)-
    \rRe T({\mib q},0)\right]
    \iIm G_0({\mib q}-{\mib k},x-\epsilon-{\rm i}\delta).
\end{eqnarray}
It is seen that $\rRe\Sigma_{1}({\mib k},\epsilon+{\rm i}\delta)$
is a function of only $\epsilon-vk$ and
$\rRe\Sigma_{2}({\mib k},\epsilon+{\rm i}\delta)$ a function of only $\epsilon$
for the dispersion given by eq. (\ref{eq:dispersion}).
Reflecting the regular behavior of $\rRe T({\mib q},0)$,
$\rRe\Sigma_{2}({\mib k},\epsilon+{\rm i}\delta)$ is a regular function
of $\epsilon$.
Actually for $|\epsilon|, |vk| \ll \epsilon_{\rm c}$, we obtain
\begin{eqnarray}
  \rRe\Sigma_{1}({\mib k},\epsilon+{\rm i}\delta)&\simeq&
    \rRe\Sigma (0,0)+c_{1}(\epsilon-vk)\\
  \rRe\Sigma_{2}({\mib k},\epsilon+{\rm i}\delta)&\simeq& c_2\epsilon,
\end{eqnarray}
where $c_{1}$ and $c_2$ are constants.
On the other hand, we note that
$\rRe\Sigma_{3}({\mib k},\epsilon+{\rm i}\delta)$
and $\iIm\Sigma({\mib k},\epsilon+{\rm i}\delta)$ are real and imaginary parts
of a function which
has a contribution
from only the low-energy region, i.e.,
$|q_x|\ll k_{\rm c}, |q_y|\ll k_{\rm c}', |x|\ll \epsilon_{\rm c}$.
\begin{eqnarray}
  \lefteqn{\rRe\Sigma_{3}({\mib k},\epsilon+{\rm i}\delta)
    +i\iIm\Sigma({\mib k},\epsilon+{\rm i}\delta)}\\
  &\simeq&
    -\int_{-\lambda}^{\lambda}\frac{{\rm d}q_x}{2\pi}
    \int_{-\lambda'}^{\lambda'}\frac{{\rm d}q_y}{2\pi}
    \int_{0}^{\epsilon}\frac{{\rm d}x}{\pi}
    \left[T({\mib q},x+{\rm i}\delta)+U_{\rm eff}\right]
    \iIm G_0({\mib q}-{\mib k},x-\epsilon-{\rm i}\delta),
\end{eqnarray}
where the cut-offs $\lambda$ and $\lambda'$ are small values compared to
$k_{\rm c}$ and $k_{\rm c}'$, respectively,
and $U_{\rm eff}\equiv 
-\lim_{q\rightarrow 0}\left[\lim_{x\rightarrow 0}T({\mib q},x+{\rm i}\delta)
\right]$.
We introduce $\Sigma_{\rm s}({\mib k},\epsilon+{\rm i}\delta)$ by
subtracting the on-shell value of
$\rRe\Sigma_{3}({\mib k},\epsilon+{\rm i}\delta)$,
\begin{equation}
  \Sigma_{\rm s}({\mib k},\epsilon+{\rm i}\delta)\equiv
    [\rRe\Sigma_{3}({\mib k},\epsilon+{\rm i}\delta)-
    \rRe\Sigma_{3}({\mib k},vk+{\rm i}\delta)]+
    i\iIm\Sigma({\mib k},\epsilon+{\rm i}\delta).
\end{equation} 
$\rRe\Sigma_{\rm s}({\mib k},\epsilon+{\rm i}\delta)$ is of interest to us, while
$\rRe\Sigma({\mib k},\epsilon+{\rm i}\delta)-
\rRe\Sigma_{\rm s}({\mib k},\epsilon+{\rm i}\delta)
=\rRe\Sigma_1({\mib k},\epsilon+{\rm i}\delta)
+\rRe\Sigma_2({\mib k},\epsilon+{\rm i}\delta)
+\rRe\Sigma_3({\mib k},vk+{\rm i}\delta)$
is related to various renormalizations such as
the shift of the chemical potential, the renormalization of the Fermi velocity,
the effects of which are taken into account by replacing
$v_0$ and $A_0$ in
eq. (\ref{eq:dispersion}) and the renormalization factor
$Z_0$ of the Green function $G_0({\mib k},\epsilon+{\rm i}\delta)$.
Then, we obtain the renormalized self-energy
$\Sigma^*({\mib k},\epsilon+{\rm i}\delta)$ and the renormalized Green function
$G^*({\mib k},\epsilon+{\rm i}\delta)$ as
\begin{eqnarray}
  \Sigma^*({\mib k},\epsilon+{\rm i}\delta)&\equiv&
    \left[\Sigma_{\rm s}({\mib k},\epsilon+{\rm i}\delta)
    \right]_{G_0\rightarrow G_0^*}
  \label{eq:gogo}\\
  G^*({\mib k},\epsilon+{\rm i}\delta)&\equiv&\frac{Z_0}{(\epsilon-vk)-
   \Sigma^*({\mib k},\epsilon+{\rm i}\delta)},
\end{eqnarray}
where $G_0^*({\mib k},\epsilon+{\rm i}\delta)\equiv Z_0/(\epsilon-\xi_k^*)$
and $\xi_k^*\equiv vk_x+Ak_y^2/2$.
The subscript $G_0\rightarrow G_0^*$ 
on the right hand side (r.h.s.) of eq. (\ref{eq:gogo}) indicates that
the calculation of $\Sigma_{\rm s}({\mib k},\epsilon+{\rm i}\delta)$ should be
performed using the renormalized Green function,
$G_0^*({\mib k},\epsilon+{\rm i}\delta)$.
We note that the on-shell value of
$\rRe\Sigma_{3}({\mib k},\epsilon+{\rm i}\delta)$,
$\rRe\Sigma_{3}({\mib k},vk+{\rm i}\delta)$,
is of the order of 
$(vk)^{2}$, and that this term gives the renormalization of a quasi-particle
energy only of the order of $k^2$.

Next we need to estimate $\Sigma^*({\mib k},\epsilon+{\rm i}\delta)$.
Under the conditions of $|q_{x}|\ll k_{\rm c},|q_{y}|\ll k_{\rm c}',
|x|\ll \epsilon_{\rm c}$,
the asymptotic form of $K^*(\mib{q},x+{\rm i}\delta)-K_{0}^*$ is given by
\begin{eqnarray}
  K^*(\mib{q},x+{\rm i}\delta)-K^*_{0}\simeq\left\{
  \begin{array}{ll}
    \displaystyle{\frac{{\rm i}Z_0^2x}{4\pi A^{1/2}v(x-vq_{x}-Aq_{y}^{2}/4)
      ^{1/2}}}&
      \mbox{$(x>vq_{x}+Aq_{y}^{2}/4)$}\\
    \displaystyle{\frac{Z_0^2x}{4\pi A^{1/2}v(-x+vq_{x}+Aq_{y}^{2}/4)
      ^{1/2}}}&
      \mbox{$(x<vq_{x}+Aq_{y}^{2}/4)$,}
  \end{array}\right. 
\end{eqnarray}
where the particle-particle correlation function $K^*(\mib{q},x+{\rm i}\delta)$
is calculated using the renormalized Green function
$G_0^*(\mib{k},\epsilon+{\rm i}\delta)$.
Using these expressions,
we obtain the asymptotic form of
$\rRe\Sigma^*({\mib k},\epsilon+{\rm i}\delta)$ in the limit of
$|\epsilon-vk|\ll A^{-1}U_{\rm eff}^2Z_0^4k^{2}$
for the case of $\epsilon>vk$, for example, as
\begin{eqnarray}
  \lefteqn{\rRe\Sigma^*({\mib k},\epsilon+{\rm i}\delta)}&&\nonumber\\
  &\simeq&
    -\frac{Z_0}{4\pi^{2}v}\int_{-\lambda'}^{\lambda'}{\rm d}q_{y}
    \int_{0}^{\epsilon}{\rm d}x
    \rRe\left[\frac{-U_{\rm eff}}{1+\dfrac{{\rm i}U_{\rm eff}Z_0^2x}
    {4\pi A^{1/2}v(Aq_{y}^{2}/4+\epsilon-vk)^{1/2}}}-
    \frac{-U_{\rm eff}}{1+\dfrac{{\rm i}U_{\rm eff}Z_0^2x}
    {4\pi A^{1/2}v(Aq_{y}^{2}/4)^{1/2}}}\right]\nonumber\\
  &\simeq&(\epsilon-vk)\frac{2}{\pi Z_0}\int_{0}^{\infty}{\rm d}q_{y}
    \left[\left(\frac{q_{y}^{2}}{4}+1\right)^{1/2}\arctan
    \left(\frac{U_{\rm eff}Z_0^2\epsilon}{4\pi A^{1/2}v(\epsilon-vk)^{1/2}
    \left(q_{y}^{2}/4+1\right)^{1/2}}\right)\right.\nonumber\\
  & &\makebox[2cm]{}
    -\left.\left(\frac{q_{y}^{2}}{4}\right)^{1/2}\arctan
    \left(\frac{U_{\rm eff}Z_0^2\epsilon}{4\pi A^{1/2}v(\epsilon-vk)^{1/2}
    \left(q_{y}^{2}/4\right)^{1/2}}\right)\right]\nonumber\\
  &\simeq&\sign[k]\frac{(\epsilon-vk)}{Z_0}\int_{0}^{\nu}{\rm d}q_{y}
    \left[\left(\frac{q_{y}^{2}}{4}+1\right)^{1/2}-
    \left(\frac{q_{y}^{2}}{4}\right)^{1/2}\right]\nonumber\\
  &\simeq&\sign[k]\frac{(\epsilon-vk)}{Z_0}\log\nu,
  \label{eq:log}
\end{eqnarray}
where
\begin{equation}
  \nu\simeq \frac{U_{\rm eff}Z_0^2|k|}{A^{1/2}(\epsilon-vk)^{1/2}}\gg 1.
\end{equation}
The calculations for the case of $\epsilon<vk$ are similar
and we obtain the final result in the limit of
$|\epsilon-vk|\ll A^{-1}U_{\rm eff}^2Z_0^4k^{2}$ as
\begin{equation}
  \rRe\Sigma^*(k,\epsilon+{\rm i}\delta)\simeq
      c\frac{(\epsilon-vk)}{Z_0}\log\frac{A|\epsilon-vk|}
      {U_{\rm eff}^2Z_0^4k^{2}},
\end{equation}
where
\begin{equation}
  c=\left\{\begin{array}{ll}
    \displaystyle{-\frac12}&\mbox{$(\epsilon>vk, k>0)$}\\
    \displaystyle{0}&\mbox{$(\epsilon<vk, k>0)$}\\
    \displaystyle{\frac12}&\mbox{$(\epsilon>vk, k<0)$}\\
    \displaystyle{1}&\mbox{$(\epsilon<vk, k<0)$}
  \end{array}.\right.
  \label{eq:coeff}
\end{equation}
Reflecting the existence of a logarithmic singularity,
the renormalization factor
$Z=\lim_{\epsilon\rightarrow vk}(1-\partial\rRe\Sigma/\partial\epsilon)^{-1}$
exhibits an anomalous behavior, $Z=0$,
except in the limit of $\epsilon\rightarrow vk-0$ in the case of $vk>0$. 

In addition to this, we have evaluated the asymptotic form of
$\iIm\Sigma^*(k,\epsilon+{\rm i}\delta)$ in the same limit of
$|\epsilon-vk|\ll A^{-1}U_{\rm eff}^2Z_0^4k^{2}$,
which is given by
\begin{eqnarray}
  \iIm\Sigma^*(k,\epsilon+{\rm i}\delta)&\simeq&
    \iIm\Sigma^*(k,vk+{\rm i}\delta)\nonumber\\
  & &+\frac{(\epsilon-vk)}{Z_0}\left[-\frac{1}{4\pi}\left(
      \log\frac{A|\epsilon-vk|}
      {U_{\rm eff}^2Z_0^4k^{2}}\right)^{2}+
      a\left(\log\frac{A|\epsilon-vk|}
      {U_{\rm eff}^2Z_0^4k^{2}}\right)+b\right]\label{eq:imss}\\
  \iIm\Sigma^*(k,vk+{\rm i}\delta)&\simeq&
    \frac{U_{\rm eff}^2Z_0^4k^2}{A}\log
    \frac{U_{\rm eff}Z_0^2|k|}{A^{1/2}\epsilon_{\rm c}^{1/2}}\equiv \Gamma_k,
\end{eqnarray} 
where $a$ and $b$ are constants which are different
in each region of $\epsilon>vk$ and $\epsilon<vk$.
Especially for $\epsilon\rightarrow vk$, only the first term on
the r.h.s. of eq. (\ref{eq:imss}) survives.
So far, the real and imaginary parts of the self-energy have been examined,
but $\Sigma(k,\epsilon+{\rm i}\delta)$ must be an analytic function of 
$\epsilon$. Actually, $\Sigma^*(k,z)$ is given as follows
for a complex variable $z$ ($\iIm\;z>0$)
in the limit of $|z-vk|\ll A^{-1}U_{\rm eff}^2Z_0^4k^{2}$.
\begin{eqnarray}
  \Sigma^*(k,z)&\simeq&
    {\rm i}\iIm\Sigma^*(k,vk+{\rm i}\delta)\nonumber\\
  & &+\frac{(z-vk)}{Z_0}\left[-\frac{\rm i}{4\pi}\left(
      \log\frac{A(z-vk)}
      {U_{\rm eff}^2Z_0^4k^2}\right)^{2}+
      a'\left(\log\frac{A(z-vk)}
      {U_{\rm eff}^2Z_0^4k^2}\right)+b'\right],
  \label{eq:sigma}
\end{eqnarray}
where $a'$ and $b'$ are complex constants related to $a$ and $b$
in eq. (\ref{eq:imss}).
From the analyticity condition, $\Sigma^*(k,\epsilon+{\rm i}\delta)$
is interrelated from $\epsilon>vk$ to $\epsilon<vk$
through the upper-half complex plane of $\epsilon$
(Fig. \ref{fig:dispersion}(c)), and therefore
the constants $a'$ and $b'$ in each region
are mutually related by a 
simple relation. 
When we see $\Sigma^*(k,z)$ as one analytic function of $z$,
the logarithmic term in the real part is related to
the second term of the r.h.s. of eq. (\ref{eq:imss}) in the imaginary part.
Actually, we obtain $\rRe\; a'=-1/2\;(1/2)$ for $vk>0\; (vk<0)$.
The leading term of the imaginary part when $\epsilon\simeq vk$, however,
is not the second term in eq. (\ref{eq:imss}) but the first one.
On the other hand, in the case of $k=k_{\rm F}$,
$\iIm\Sigma(k_{\rm F},\epsilon+{\rm i}\delta)\propto\epsilon^2\log\epsilon$ was
obtained,\cite{bib:Hodges}
which corresponds to the first term on the r.h.s. of eq. (\ref{eq:imss}),
and was the basis of the claim of the Fermi-liquid state.
As has been demonstrated, however,
the present logarithmic term of the real part cannot be deduced
from this imaginary part using the Kramers-Kronig transformation.
This has not been noted in previous studies. 

Our new result of a logarithmic singular behavior 
in the real part of the self-energy implies
that the renormalization factor $Z$ vanishes in the low-energy limit.
 
In addition, we note the relationship between the results in the
{\em t}-matrix approximation and the second-order perturbation theory.
When we focus only on the forward scattering process,
the self-energy as an analytic function of complex variable $z$ is obtained
by the second-order perturbation theory as
\cite{bib:Ogata}
\begin{equation}
  \Sigma^*(k,z)\propto
    {\rm i}U^2z^2\log\frac{(z-vk)}{\epsilon_{\rm c}},
\end{equation}
and there is no logarithmic term in the real part of the self-energy.
In the case of the {\em t}-matrix approximation,
we found that the self-energy is 
a logarithmic function not only of $\epsilon-vk$ but
also of $U_{\rm eff}$, as shown in eq. (\ref{eq:sigma}).
Hence, the logarithmic term 
in the real part of the self-energy
obtained by the {\em t}-matrix approximation is a non-perturbative effect
of $U$, which implies that $U=0$ can be a singular point in two dimensions.

Investigating the derivation in eq. (\ref{eq:log}) of the logarithmic term in
the real part of the self-energy near the on-shell region,
we can see that the logarithmic term is related to the property of the
{\em t}-matrix in the region of 
$|x-vq_x|< A^{-1}U_{\rm eff}^2Z_0^4k^{2}, 
Aq_y^2< A^{-1}U_{\rm eff}^2Z_0^4k^{2}, 0<|x|<|\epsilon|$.
This indicates that the on-shell electrons are responsible to
the logarithmic term of the self-energy.

Anderson proposed that the special feature of the forward scattering
process leads to the vanishing renormalization factor, $Z=0$,
i.e., the breakdown of the Fermi-liquid, and our results
have confirmed this based on the conventional many-body perturbation
theory using the Feynman diagrams.

So far the calculations of the self-energy have been carried out
based on the Green function without damping, $\Gamma_k$.
The resulting logarithmic behavior in the real part of the self-energy,
however, has been seen only in the energy region
$|\epsilon-vk|\ll A^{-1}U_{\rm eff}^2Z_0^4k^{2}\Gamma_k$,
implying the importance of the self-consistent treatment.
Actually, $\Gamma_k$ results in finite $Z$
even when $X\rightarrow 0$, i.e., $G(k,\epsilon+{\rm i}\delta)\simeq
Z_0/[a(X)X+{\rm i}\Gamma_k]$ with $a(0)\neq 0$. However, $a(X)$ will have a
strong dependence on $X=\epsilon-vk$ as $a(X)=a_0[1+O(|X|/\Gamma_k)]$,
which is different from a conventional Fermi-liquid.

K.Y. acknowledges Hiroshi Kohno for valuable discussions.
K.Y. also thanks
JSPS Research Fellowships for Young Scientists.
This work is financially supported by a Grant-in-Aid for
Scientific Research on Priority Area "Anomalous Metallic State near
the Mott Transition" (07237102) from the Ministry of Education,
Science, Sports and Culture.

\end{document}